\documentclass[A4paper,preprint,12pt,tightenlines,amsmath,amssymb,floatfix,nofootinbib,aps,raggedbottom]{revtex4-2}

\usepackage{graphicx}
\usepackage{bm}
\usepackage{url,multirow,xcolor}
\usepackage[font=footnotesize,margin=24pt,justification=centerlast]{caption}

\begin{document}

\title{ Investigation of the cause of the discrepancies between calculated running sums for nuclear matrix elements of two-neutrino double-$\bm{\beta}$ decay \\ \vspace{5pt} }

\author{J.\ Terasaki \\ \vspace{0pt}}
\affiliation{ Institute of Experimental and Applied Physics\hbox{,} Czech Technical University in Prague, Husova 240/5, 110\hspace{3pt}00 Prague 1, Czech Republic}

\affiliation{}

\begin{abstract} 
A qualitative difference in the running sum for the nuclear matrix element of the two-neutrino double-$\beta$ decay of $^{136}$Xe has been found four years ago between the quasiparticle random-phase approximation (QRPA) and shell model calculations. The former result has large increase and decrease with respect to the excitation energy of the intermediate state, and the latter one is an almost monotonically and mildly increasing function. My QRPA calculations independent of the above one do not have a remarkable decrease. This discrepancy is a serious problem affecting the reliability of calculations of the neutrinoless double-$\beta$ decay, and the cause was unknown. I perform several relevant test calculations and consider analytically to find the cause, which is found to be in the strength of the attractive interactions. The possible major local decrease in the running sum is also explained analytically. The  interactions of my QRPA calculation are appropriate in terms of the strength, thus, the almost monotonic behavior is reasonable. 
\end{abstract}

%
\maketitle
\newpage
\section{\label{sec:introduction}Introduction}
Studies of the nuclear matrix element (NME) of the double-$\beta$ decays continue actively by many researchers to obtain reliable prediction of the NME of the neutrinoless double-$\beta$ ($0\nu\beta\beta$) decay. This NME is necessary for designing new detectors to observe that extremely rare decay \cite{Eji02}, if it occurs, which is a key phenomenon consolidating the foundation of theories \cite{Fuk86} to explain the matter-antimatter imbalance in the current universe because the Majorana (self conjugate) neutrino is necessary for those theories beyond the standard model. If this decay is observed with sufficient statistics, the effective neutrino mass (Majorana mass) can be determined from the measured decay probability and the calculated $0\nu\beta\beta$ NME \cite{Fur39}. The effective neutrino mass is a mass scale parameter of the neutrino, which has been unknown for a long time, and that effective mass gives a constraint to the Pontecorvo-Maki-Nakagawa-Sakata (PMNS) matrix \cite{Pon57} including two Majorana phases for the Majorana neutrino. The determination of the PMNS matrix is one of the most important subjects in the current neutrino physics because this matrix is necessary for calculating the transition strengths of any reactions involving the neutrinos. 
Since the PMNS matrix is not yet fully determined, a reliable prediction of the $0\nu\beta\beta$ NME is necessary. 

However, a problem has been known for more than thirty years in the calculations of the $0\nu\beta\beta$ NME \cite{Eng17}; the calculated values are distributed in a range of the maximum-to-minimum ratio of 2$-$3, and this problem  causes a large uncertainty in the effective neutrino mass. The origin of this problem is in the nuclear calculations. Recently, new information related to this long-standing problem has been obtained \cite{Gan19} in the running sum for the NME of the two-neutrino double-$\beta$ ($2\nu\beta\beta$) decay of $^{136}$Xe $\rightarrow$ $^{136}$Ba with respect to the intermediate-state energy. The behavior of this  running sum is quite different depending on the calculations. It is easily speculated that if the components of the $2\nu\beta\beta$ NME are so calculation dependent, one is far from the reliable prediction of the $0\nu\beta\beta$ NME. The purpose of this article is to clarify the cause of this problem of running sum and to discuss what behavior is close to the reality. 

In Sec.~\ref{sec:variety_of_running_sums}, the running sums for the $2\nu\beta\beta$ NME of four groups are presented, and the discrepancy problem of the running sum is defined. The effects of the continuum single-particles are also examined. 
In Sec.~\ref{sec:high-j_orbitals}, the effects of high angular-momentum orbitals on the $2\nu\beta\beta$ NME are investigated in comparison of my quasiparticle random-phase approximation (QRPA) and a shell model calculation. Tests of different interactions are made in Sec.~\ref{sec:Increasing_interaction_strength}. This is an important step to clarify the cause of the problem. 
In Sec.~\ref{sec:Analytical_discussion}, I discuss analytically the behavior of the components of the $2\nu\beta\beta$ NME, and the interaction strength is shown to be the key point. Subsequently (Sec.~\ref{sec:validity_interaction}), the validity of my interaction is examined, and the reason for the sensitivity of the double-$\beta$ NME to the interaction is discussed. Section \ref{sec:Summary} is the summary of this study.

\section{\label{sec:variety_of_running_sums} Variety of running sums}

The $2\nu\beta\beta$ NME $M^{(2\nu)}$ can be approximated by the Gamow-Teller (GT) NME $M^{(2\nu)}_\textrm{GT}$;  
\begin{eqnarray}
M^{(2\nu)} \simeq M^{(2\nu)}_\textrm{GT}=  \sum_{B} \frac{m_e c^2}{E_B - \bar{M}}
\langle F| \bm{\sigma}\tau^- |B\rangle
\cdot\langle B| \bm{\sigma}\tau^-| I\rangle, \label{eq:M2v}
\end{eqnarray}
where $|I\rangle$, $|B\rangle$, and $|F\rangle$ are the initial, intermediate, and final nuclear state, respectively, and $E_B$ denotes the energy of $|B\rangle$. The GT operator is denoted by $\bm{\sigma}\tau^- $ with the spin Pauli matrices $\bm{\sigma}$ and the operator changing a neutron to a proton $\tau^-$. 
The Fermi component can be ignored under the approximation of the exact isospin symmetry. 
$\bar{M}$ is the mean value of the masses of the initial and final nuclei, and $m_e$ is the electron mass. 
In the QRPA, the $2\nu\beta\beta$ GT NME is calculated by  
\begin{eqnarray}
M^{(2\nu)}_\textrm{GT} \simeq  \sum_{B_I,B_F} \frac{m_e c^2}{E_B - \bar{M}}
\langle F| \bm{\sigma}\tau^- |B_F\rangle
\langle B_F | B_I \rangle 
\cdot\langle B_I| \bm{\sigma}\tau^-| I\rangle. \label{eq:M2v_QRPA}
\end{eqnarray}
$|B_I\rangle$ with energy $E_{BI}$ and $|B_F\rangle$ with energy $E_{BF}$ are the intermediate-nuclear states obtained by the QRPA based on $|I\rangle$ and $|F\rangle$, respectively. Usually, $E_B$ is replaced by the average of $E_{BI}$ and $E_{BF}$. It has been discussed \cite{Ter18} that if two calculations using $E_B$ = $E_{BI}$ and $E_B$ = $E_{BF}$ give close results, the QRPA is a good approximation. This applies to  $^{136}$Xe $\rightarrow$ $^{136}$Ba \cite{Ter19}, and  $E_B$ = $E_{BI}$ is  used in my calculations, unless otherwise mentioned, because $^{136}$Xe is slightly farther from the transitional region than $^{136}$Ba.  

\begin{figure}[t]
\includegraphics[width=0.5\columnwidth]{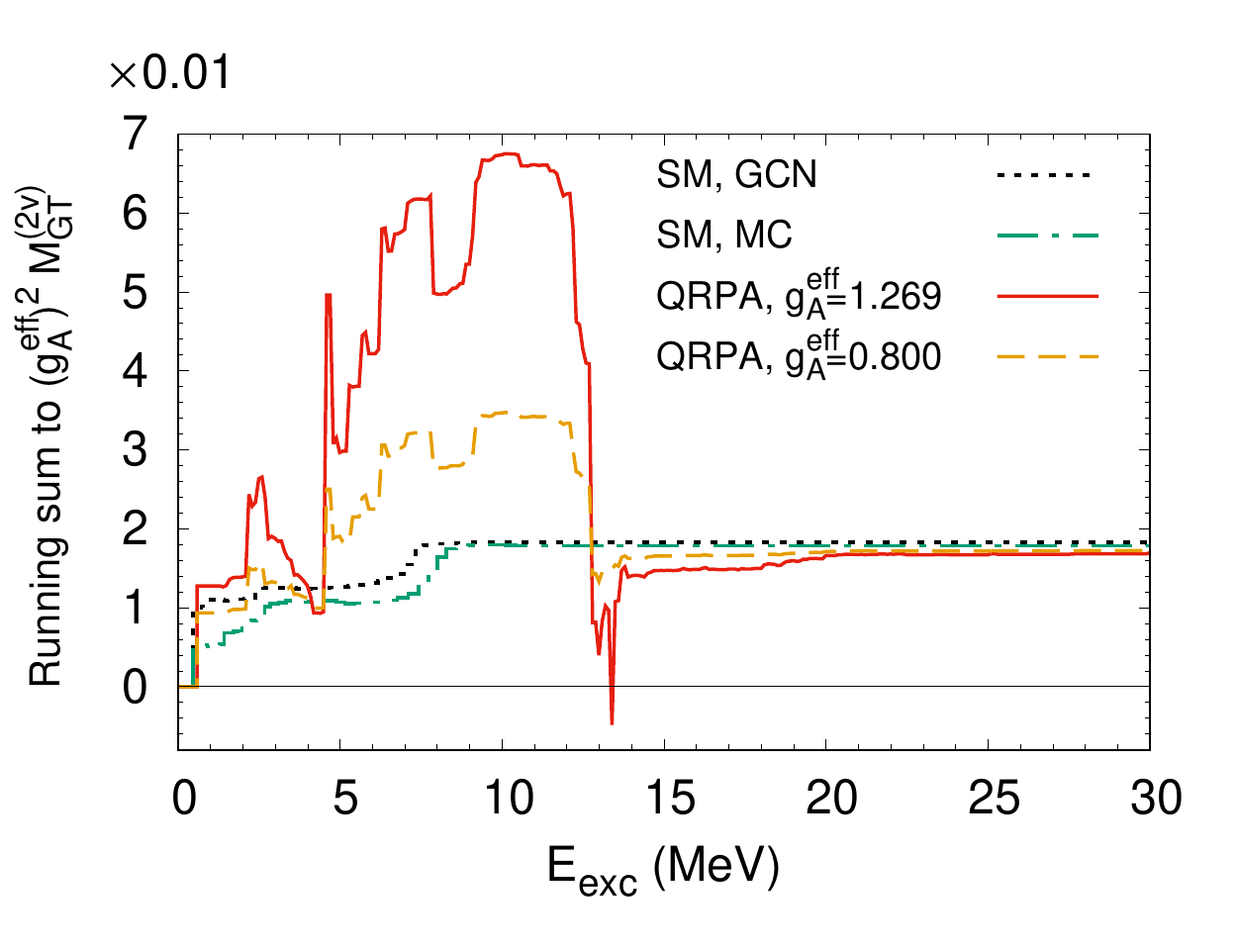}
\vspace{-10pt}
\caption{ \protect \label{fig:QRPA_SM} \baselineskip=13pt 
Running sum for $(g_A^\textrm{eff})^2 M^{(2\nu)}_\textrm{GT}$ of $^{136}$Xe $\rightarrow$ $^{136}$Ba calculated by QRPA and shell model (SM) as functions of excitation energy $E_\textrm{exc}$ of the intermediate nucleus $^{136}$Cs \cite{Gan19}. The effective axial-vector current coupling is denoted by $g_A^\textrm{eff}$. The shell model calculation uses two interactions GCN and MC, and $g_A^\textrm{eff}$ is used for fitting the experimental half-life of the $2\nu\beta\beta$ decay. The two QRPA calculations use the $G$ matrix based on the Argonne V18 nucleon-nucleon potential with different $g_A^\textrm{eff}$, and the strength of the isoscalar proton-neutron interaction is used for fitting the half-life.  
}
\end{figure}

Figure \ref{fig:QRPA_SM} shows the running sum multiplied by the effective axial-vector current coupling $g_A^\textrm{eff}$ squared\footnote{ This figure is a reuse from Ref.~\cite{Gan19}, ``Precision Analysis of the $^{136}$Xe Two-Neutrino $\beta\beta$ Spectrum in KamLAND-Zen and Its Impact on the Quenching of Nuclear Matrix Elements", DOI: https://doi.org/10.1103/PhysRevLett.122.192501. }. That of the QRPA by the group of Ref.~\cite{Sim18b} has large increase and decrease, and the result of the shell model used in a series of calculations \cite{Men09} is almost monotonically increasing toward the convergence.  I call this qualitative difference the discrepancy problem in this article. Since the components give more information than the integrated values, this new problem is a clue for improving the double-$\beta$ NME calculations.  

\begin{figure}[t]
\includegraphics[width=0.5\columnwidth]{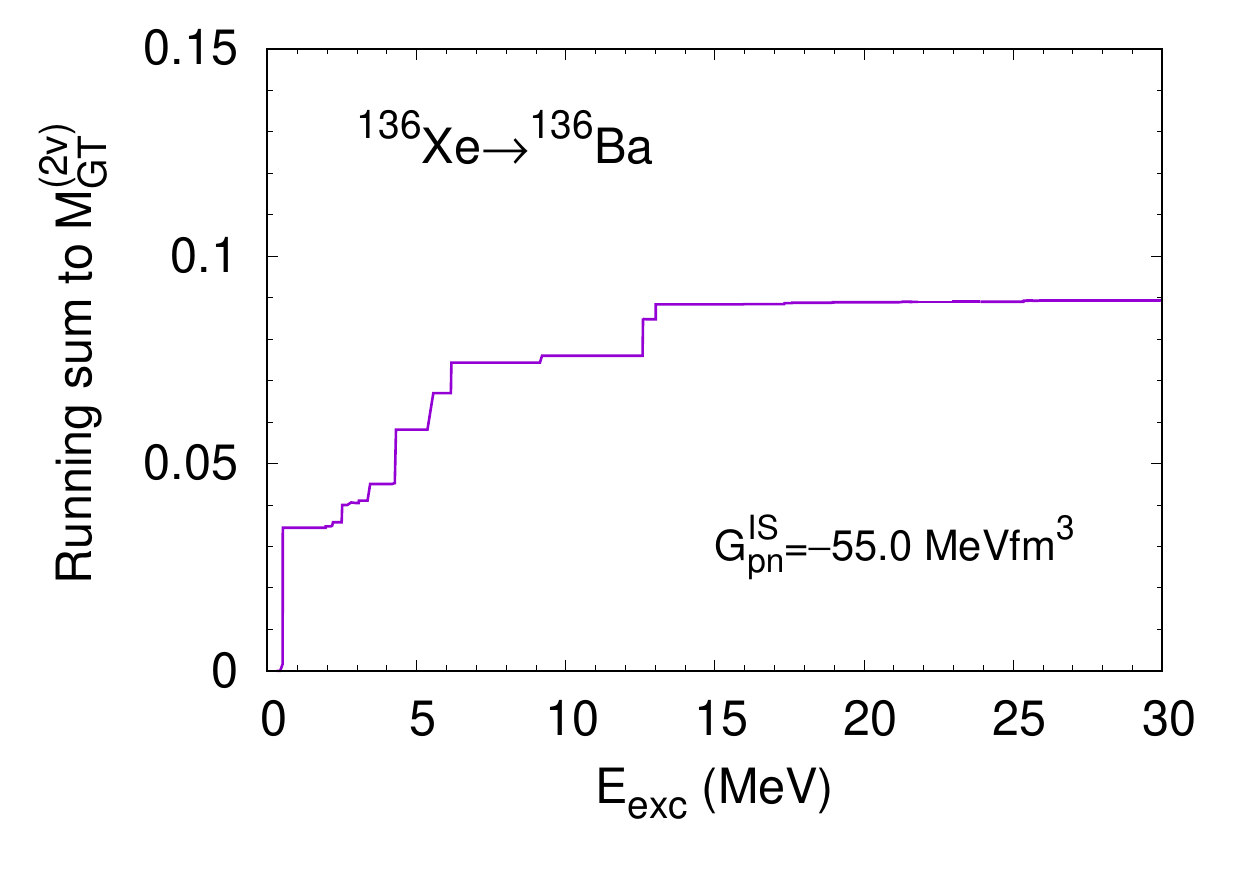}
\vspace{-10pt}
\caption{ \protect \label{fig:runsum-55.0} \baselineskip=13pt 
Running sum for $M^{(2\nu)}_\textrm{GT}$ of $^{136}$Xe $\rightarrow$ $^{136}$Ba by my QRPA calculation as a function of $E_\textrm{exc}$ of $^{136}$Cs. The Skyrme interaction (SkM$^\ast$ \cite{Bar82}) is used. The strength of the contact isoscalar proton-neutron pairing interaction $G^\textrm{IS}_{pn}$ is $-$55.0 MeVfm$^3$, which was determined according to the method of Refs.~\cite{Ter16,Ter20}.
 }
\end{figure}

There are two more calculations relevant to this problem for $^{136}$Xe. One is a shell model calculation \cite{Hor13}, which introduced $0g_{9/2}$ and $0h_{9/2}$ orbitals to their calculation in addition to the usual shells of $2s_{1/2}$, $1d_{5/2,3/2}$, $0g_{7/2}$, and $0h_{11/2}$. 
They found appreciable increase and decrease in their $M^{(2\nu)}_\textrm{GT}$, when the involvement of the two new orbitals  are extended step by step through the intermediate excited states and the configuration mixing in the initial and final states (discussed more below). They did not show the running sum, but their running sum of $^{48}$Ca shows the behavior of increase and decrease \cite{Hor07}. Thus, the discrepancy problem is not limited to $^{136}$Xe. 
The fourth example is my QRPA calculation \cite{Ter19}. As shown in Fig.~\ref{fig:runsum-55.0}, my running sum for the $M^{(2\nu)}_\textrm{GT}$ is a monotonically increasing function. Thus, two calculations have the running sum with large increase and decrease or large variations in the NME of the test  calculations, and the other two show nearly monotonic running sums. Each group has QRPA and shell model calculations. Therefore, the theoretical differences of the two methods are not the cause of that discrepancy problem. The feature of the QRPA is that large single-particle space can be used, but the creation of the excited states from the ground state is described only by two-quasiparticle additions  (forward) and annihilations (backward). The ground-state correlations are included, but it is perturbative. The shell model wave functions include more many-body correlations than the QRPA wave functions do, but the single-particle space is small compared to that used in the QRPA calculations. The shell model is a nonperturbative method. 

\begin{figure}[t]
\includegraphics[width=0.5\columnwidth]{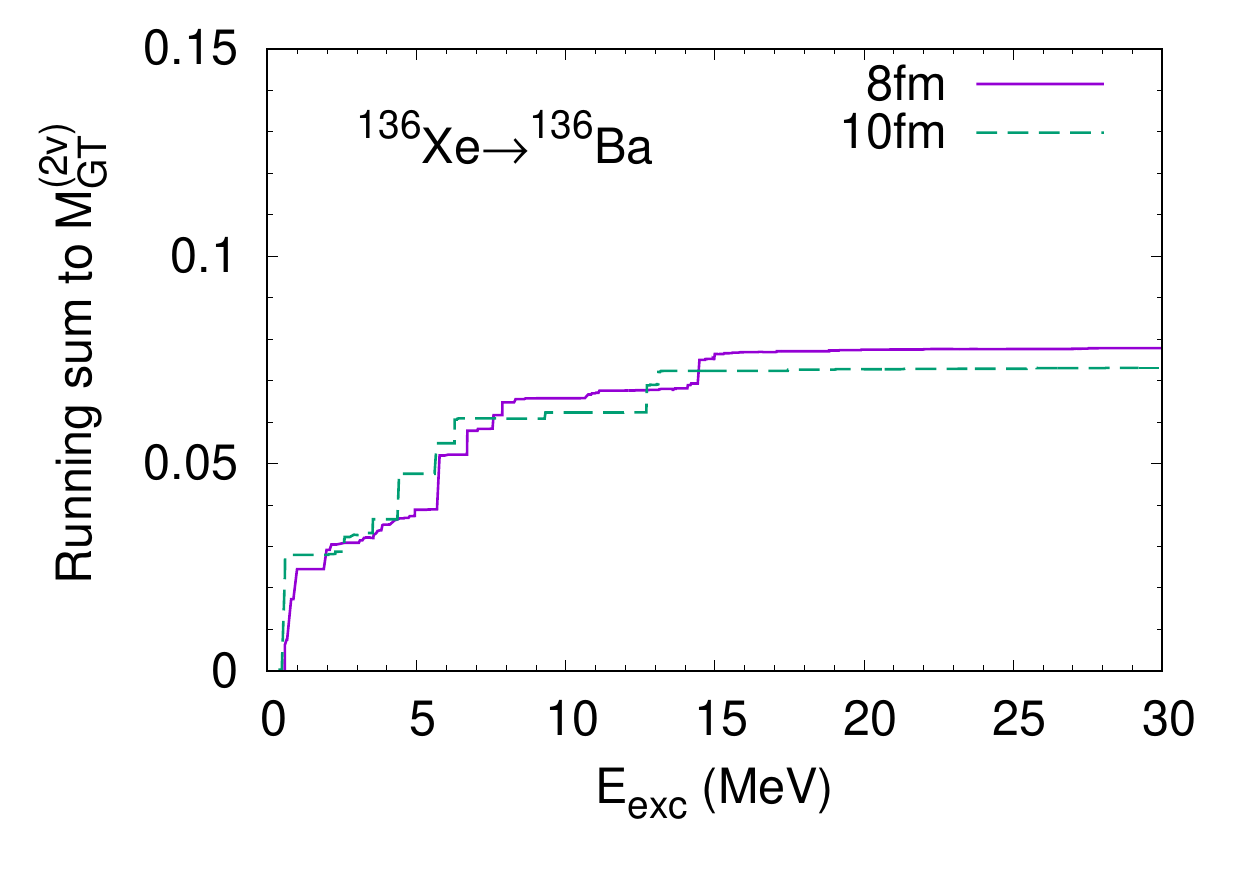}
\vspace{-10pt}
\caption{ \protect \label{fig:runsum8_10fm} \baselineskip=13pt 
The same as Fig.~\ref{fig:runsum-55.0} but for the cylinder box sizes of 10 and 8 fm for the coordinate representation of the quasiparticle wave functions in the HFB calculation. }
\end{figure}

The harmonic-oscillator basis to represent the single-particle wave functions may have unrealistic couplings between the states in the continuum and bound regions because any wave functions are spatially restricted due to the infinite wall of the harmonic-oscillator potential. 
The influence of this problem can be examined by my QRPA calculation because a large cylindrical box is used in my Hartree-Fock-Bogoliubov (HFB) calculation to obtain the quasiparticle wave functions on the two-dimensional coordinate mesh. The radius of the box in the $xy$ plane containing the circle cross section of the cylinder and half the height to the $z$ direction are both 20 fm. The root-mean-square radius of the ground state of $^{136}$Xe is 4.8 fm according to my HFB calculation using the Skyrme (the parameter set SkM$^\ast$ \cite{Bar82}) and contact pairing interactions. For the details of the HFB  calculation, see Refs.~\cite{Ter03,Bla05,Obe07}. The same interactions were used for the QRPA calculation. I performed the running-sum calculations with the 10- and 8- fm boxes, and no decreasing part was found; see Fig.~\ref{fig:runsum8_10fm}.  Thus, the harmonic-oscillator basis is not the cause of the discrepancy problem. However, it is stressed that the large box size is important for the reliable prediction of the $0\nu\beta\beta$ NME.  

\section{\label{sec:high-j_orbitals} Effects of high angular-momentum orbitals}
In a discussion in Ref.~\cite{Hor13}, the authors focused on the number of particles excited from the $0g_{9/2}$ or to the $0h_{9/2}$ orbital relative to the others.  This number for the initial and final states is denoted by $n(0^+)$, and the number for the intermediate $1^+$ states is denoted by $n(1^+)$. They performed the calculations for analysis with  the restrictions of 
$[n(0^+), n(1^+)]$ = [0,0], [0,1], [1,1], and [1,2]. 
I simulate their analysis by suppressing the forward or backward amplitudes with these two orbitals of the QRPA solutions in the NME calculation. The forward amplitudes are the main parts of the transition to the excited states, and the backward amplitudes have a role to create the ground-state correlations. 
Prior to the comparison, I mention technical uncertainties in the comparison of their shell model and my  QRPA calculations. 
I do not know the definition of their single-particle basis used in their analysis. My single-particle basis is defined by the canonical basis \cite{Rin80} of the HFB calculations. I manipulate the contributions of those orbitals of only the protons because those basis states obtained by the diagonalization of the density matrix of the system with no pairing gap  (the neutrons of $^{136}$Xe) in the $M$-scheme have mixing of different angular momenta due to the degeneracy of the occupation probability. But the contributions of relevant neutrons should be automatically suppressed in the charge change transformation. 

Their result and my corresponding one are compared in Table \ref{tab:ourQRPA_HoroiSM}. 
Two differences are clearly seen. One is that the variation of my results is much smaller than the corresponding variations of their results. The maximum variation of my QRPA results is only 5 \%, and that of the shell model is a factor of 4.5. In fact, the shell model result increases by $n(1^+)$ = 1 (from Test ID 4 to 3), and the NME decreases drastically by the extension of the ground-state correlations (Test ID 1) and further by the higher-order configurations of the excited 1$^+$ state (Test ID 5).  The other difference is that my NME values are much larger than the shell model values overall except for $[n(0^+),n(1^+)]$ = [0,1]. The relative largeness of the QRPA values is the usual tendency between the shell model and the QRPA calculations \cite{Eng17}, however, the difference between their most extended result and the QRPA values is more profound than ever known. 
If a general property of the $0g_{9/2}$ and $0h_{9/2}$, e.g.,~the high angular momentum, is the cause of their large variation, a similar effect should also appear in my calculation; thus, the cause  is something else. This comparison indicates a possibility that if the same single-particle space is used in the QRPA and shell model calculations, the discrepancy problem remains. 
\begin{table}
\caption{\label{tab:ourQRPA_HoroiSM} $M^{(2\nu)}_\textrm{GT}$ of my QRPA and corresponding $M^{(2\nu)}_\textrm{GT}$ of shell model calculations \cite{Hor13} for $^{136}$Xe with various restrictions on contributions of $0g_{9/2}$ and $0h_{9/2}$. The QRPA calculation has the Fermi component, which is only a few percent of $M^{(2\nu)}_\textrm{GT}$ and omitted. Each row indicates the corresponding calculations of the two methods, if available. For the definition of $n(0^+)$ and $n(1^+)$ see text. Quenching factor for the GT operator is not used. }
\begin{ruledtabular}
\begin{tabular}{ccccc}
Test ID & \multicolumn{2}{c}{My QRPA calculation} & \multicolumn{2}{c}{Shell model calculation \cite{Hor13}} \\
\cline{2-3} \cline{4-5} \\[-11pt]
 & Suppressed & $M^{(2\nu)}_\textrm{GT}$ & $[n(0^+),n(1^+)]$ & $M^{(2\nu)}_\textrm{GT}$ \\[-2pt]
 & amplitudes  & & & \\[2pt]
\hline\\[-10pt]
1  &  None &  0.0874 & [1,1] & 0.0345 \\
2  &  Forward & 0.0888 &         &       \\
3  &  Backward  & 0.0918 & [0,1] & 0.0849 \\
4  &  Both  & 0.0932 & [0,0] & 0.0579 \\
5  &    &   &  [1,2] & 0.0187 
\end{tabular}
\end{ruledtabular}
\end{table}
%

\section{\label{sec:Increasing_interaction_strength} Increasing interaction strength}
\begin{figure}[t]
\includegraphics[width=0.5\columnwidth]{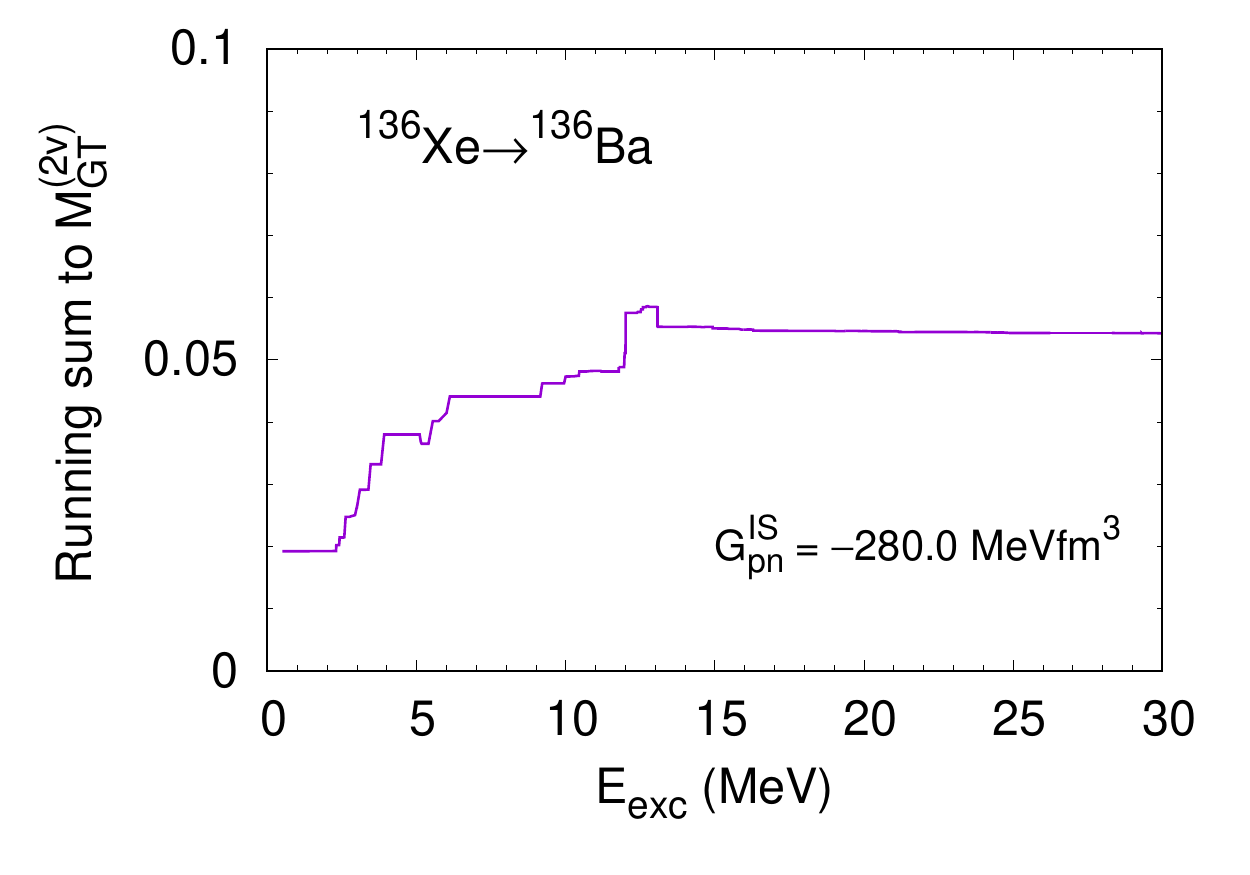}
\vspace{-10pt}
\caption{ \protect \label{fig:runsum-280.0} \baselineskip=13pt 
The same as Fig.~\ref{fig:runsum-55.0} but for $G^\textrm{IS}_{pn}=-280.0$ MeVfm$^3$ and $E_B$ = $E_{BF}$ [see Eq.~(\ref{eq:M2v_QRPA})]. }
\end{figure}

One of the two normalized running sums of the QRPA calculation in Fig.~\ref{fig:QRPA_SM} shows much larger increase and decrease than the other. If the factor of $(g_A^\textrm{eff})^2$ is removed from the result, the QRPA $M^{(2\nu)}_\textrm{GT}$ with ``$g_A^\textrm{eff}=1.269$" is 40 \% of that of the ``$g_A^\textrm{eff}=0.800$" calculation.  
In this comparison, the significant increase at the excitation energy of the intermediate state $E_\textrm{exc}$ = 4.5 MeV and the largest decrease at $E_\textrm{exc}$ = 12.5 MeV of the former calculation are comparable with those of the latter  calculation.  The lower value of the NME can be speculated to reflect on the stronger attractive residual interaction because the known systematic difference between the double-$\beta$ NMEs of the QRPA and the shell model mentioned above indicates that the attractive many-body correlations lower the $0\nu\beta\beta$ NME. 
Thus, it is worthy making a test of increasing the attractive interaction strength. The simplest method for this test is to enhance the strength of the isoscalar proton-neutron ($pn$) pairing interaction $G^\textrm{IS}_{pn}$ because this interaction effectively affects the GT transition strength.
 Figure \ref{fig:runsum-280.0} shows my running sum with $G^\textrm{IS}_{pn}=-280.0$ MeVfm$^3$. 
It is emphasized that this is a very large value close to the breaking point of the QRPA. 
In fact, $E_B$ = $E_{BF}$ was used in Eq.~(\ref{eq:M2v_QRPA}) for this calculation. With $E_B$ = $E_{BI}$ the small decrease at $E_\textrm{exc}$ = 12.9 MeV, which is the peak energy of the GT giant resonance, was not obtained. It is an important clue that a decrease was found in my calculation because this indicates the possibility that the strong attractive interaction caused this decrease.  The NME of this test is 61 \% of my normal result. This decrease is also an effect of the strong attractive interaction. 

\begin{figure}[t]
\includegraphics[width=0.5\columnwidth]{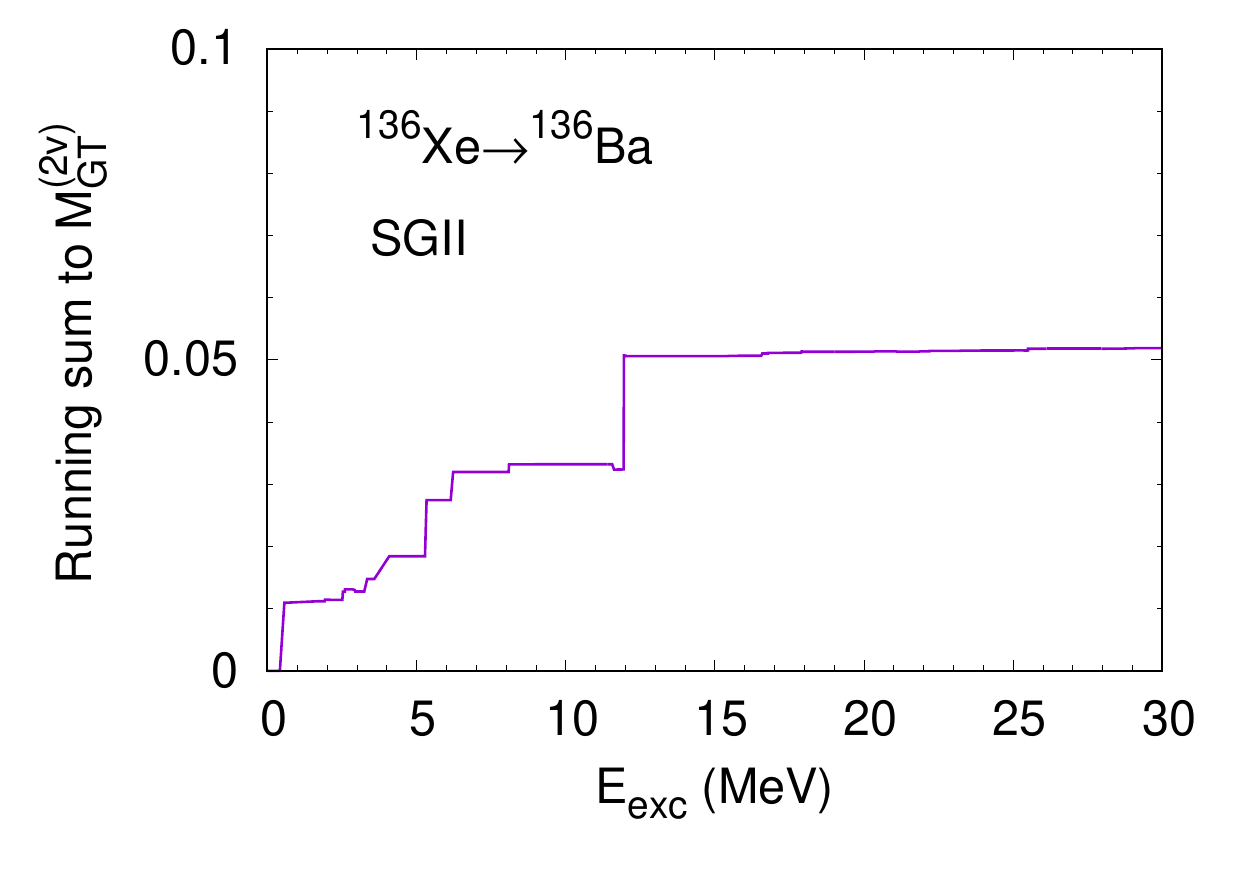}
\vspace{-10pt}
\caption{ \protect \label{fig:runsum_SGII} \baselineskip=13pt 
The same as Fig.~\ref{fig:runsum-55.0} but for the Skyrme parameter set SGII \cite{Gia81} and $G^\textrm{IS}_{pn}$ = $-$36.2 MeVfm$^3$, which was determined by the method of Refs.~\cite{Ter16,Ter20}. }
\end{figure}

Next, I discuss an influence of the particle-hole interactions. I performed another QRPA calculation with the Skyrme parameter set SGII \cite{Gia81}, which is one of the often used parameter sets for studies of the GT strength functions. The running sum is illustrated in Fig.~\ref{fig:runsum_SGII}. The contribution of the giant resonance at $E_\textrm{exc}$ = 12 MeV is much larger than that of the SkM$^\ast$ calculation (see Fig.~\ref{fig:runsum-55.0}), and the $M^{(2\nu)}_\textrm{GT}$ is 57 \% of that with SkM$^\ast$. The  SGII value is close to that with the extremely enhanced $G^\textrm{IS}_{pn}$ of $-280.0$ MeVfm$^3$ and SkM$^{\ast}$; see Fig.~\ref{fig:runsum-280.0}. In fact, the binding energy per nucleon $E_b/A$ of $^{136}$Xe of the HFB ground state (SGII) is 8.603 MeV,  and the experimental value is 8.396 MeV \cite{nndc}. The $E_b/A$ (SkM$^\ast$) is 8.415 MeV, of which the deviation from the experimental value is one order of magnitude smaller than that of SGII. The large overbinding of SGII and the lower $M^{(2\nu)}_\textrm{GT}$ indicate that  this NME is lowered by the attractive correlations between nucleons. Because of the $E_b/A$, the NME of SkM$^{\ast}$ is more reliable than that of SGII.  No major decreasing behavior was found in the running sum with SGII. 

\section{\label{sec:Analytical_discussion} Analytical discussion}
So far four possibilities of the cause or clues on the discrepancy problem of the running sum were examined.
Let me turn to analytical discussion on how the decrease in the running sum, implying a negative component of the NME, can be explained. My intention is to clarify whether the condition to cause this behavior is one of the items examined above. An analytical discussion of the (Q)RPA is enabled by assuming that the interaction is given in the form of a product of two one-body operators (separable interaction) and ignoring the exchange terms in the  derivation of the dynamical equation \cite{Rin80}. There is a method to derive the separable interactions from any general two-body interaction \cite{Duf96}. The ignorance of the exchange terms may be compensated by a renormalization of the interaction strength. Thus, those conditions are not unrealistic.  
The operator to create an intermediate state by acting on the initial state is set in the RPA to 
\begin{eqnarray}
O^\dagger_\Lambda = \sum_{\mu i}(\psi_{\mu i} c^\dagger_\mu c_i -\varphi_{\mu i} c^\dagger_i c_\mu).
\end{eqnarray}
The label of the excited state is $\Lambda$. The forward ($\psi_{\mu i}$) and backward ($\varphi_{\mu i}$) amplitudes depend on $\Lambda$, but this is omitted. I consider a single-charge transition from a neutron (roman letter is used for the state label) to a proton (greek letter is used); the creation (annihilation) operators of them are denoted by $c^\dagger_\mu$ and $c^\dagger_i $ ($c_\mu$ and $c_i $). 
The pairing correlations are ignored for simplicity in this discussion. 
The matrix element of the two-body interaction is written 
\begin{eqnarray}
V_{\mu j, i\nu}=\frac{1}{2}\chi C_{\mu i} C_{\nu j}, 
\end{eqnarray}
where $\chi$ is a negative interaction strength, and $C_{\mu i}$ denotes the matrix element of a single-charge change operator. 
The secular equation to determine the RPA eigen energy $E$ can be obtained as 
\begin{eqnarray}
S(E)  \equiv 2 \sum_{\mu i} \frac{ |C_{\mu i}|^2 (\epsilon_\mu - \epsilon_i) }{ (\epsilon_\mu - \epsilon_i)^2 - E^2 } = -\frac{1}{\chi}. \label{eq:secular}
\end{eqnarray}
$E$ is a function of $\Lambda$, but it is omitted. The single-particle energies $\epsilon_\mu$ and $\epsilon_i$ are used. The amplitudes are also obtained analytically;
\begin{eqnarray}
\psi_{\mu i} = \frac{ NC_{\mu i} }{ \epsilon_\mu - \epsilon_i - E }, \hspace{10pt}
\varphi_{\mu i} = \frac{ NC_{\mu i} }{ \epsilon_\mu - \epsilon_i + E }. \label{eq:amp}
\end{eqnarray}
$N$ is the normalization factor 
(suffix $\Lambda$ is omitted)
determined by 
\begin{eqnarray}
\sum_{\mu i}( \psi_{\mu i}^2 - \varphi_{\mu i}^2 ) = 1.
\label{eq:normalization}
\end{eqnarray}
The arbitrary constant phase of $|B\rangle$ does not affect the double-$\beta$ NME, thus I set $N$ positive for all $\Lambda$ to simplify discussion. 
Equations (\ref{eq:secular}) and (\ref{eq:amp}) are not a perturbative expansion. 

If $\chi=0$, I have $E=\epsilon_\mu - \epsilon_i$ and $\psi_{\mu i} =1$; the other amplitudes vanish. The $\beta$ decay does not occur for the candidates of the $0\nu\beta\beta$ decay used in the experiments, thus I  can assume $\epsilon_\mu - \epsilon_i > 0$. If weak $\chi < 0$ is introduced, $E$ is lowered
 [see Eq.~(\ref{eq:secular})], and simultaneously other minor but many amplitudes appear.  
Because of the normalization condition, the major $\psi_{\mu i}$ decreases as 
\begin{eqnarray}
\psi_{\mu i} \simeq
1-\frac{1}{2}\sum_{\mu^\prime i^\prime \neq \mu i}
\frac{ C_{\mu^\prime i^\prime}^2 }{ C_{\mu i}^2 }
\left( \frac{\epsilon_\mu-\epsilon_i-E}{\epsilon_{\mu^\prime}-\epsilon_{i^\prime}-E} \right)^2. 
\label{eq:decrease_psi}
\end{eqnarray}
This decrease occurs in many RPA solutions affected by the interaction, thus, the running sum is lowered nearly overall. 
 
I consider an effect of enhancement of $\chi$ taking into account the next higher $\epsilon_{\mu^\prime} - \epsilon_{i^\prime}$ to $\epsilon_\mu-\epsilon_i$. Another $E$ originally close to $\epsilon_{\mu^\prime} - \epsilon_{i^\prime}$ is possible to be closer to $\epsilon_\mu - \epsilon_i$, and the $\psi_{\mu i}$ with this $E$ has the opposite sign to the previous one. This is the origin of the negative component of the double-$\beta$ NME. An illustrative example is shown in Appendix \ref{sec:appendix1}. 
The possibility of the sign inversion of the  GT NME components is larger for those with the major component $\mu i$ being the spin-orbit partner than the others because $S(E)$ needs positive terms to satisfy Eq.~(\ref{eq:secular}). 
I analyzed my QRPA solution based on $^{136}$Xe corresponding to the giant resonance peak of the GT$^-$ ($n$ $\rightarrow$ $p$) strength function. The canonical-quasiparticle basis \cite{Rin80} is used for representing $O^\dagger_\Lambda$ in my QRPA calculation. The largest components of $O^\dagger_\Lambda$ of my normal calculation were found to be $\psi_{\mu i}$ $\simeq$ 0.2 with $\mu=p0h_{9/2}$ and a few $i$'s including $n0h_{11/2}$.  For the calculation with $G^\textrm{IS}_{pn}$ = $-$280 MeVfm$^3$, I found ten $\psi_{\mu i}$ $\simeq$ 0.2 of this configuration in the QRPA solution of the GT giant resonance peak, which has the negative contribution to the $M^{(2\nu)}_\textrm{GT}$ of Fig.~\ref{fig:runsum-280.0}. My numerical result is consistent with this analytical consideration. 

\begin{figure}[t]
\includegraphics[width=0.5\columnwidth]{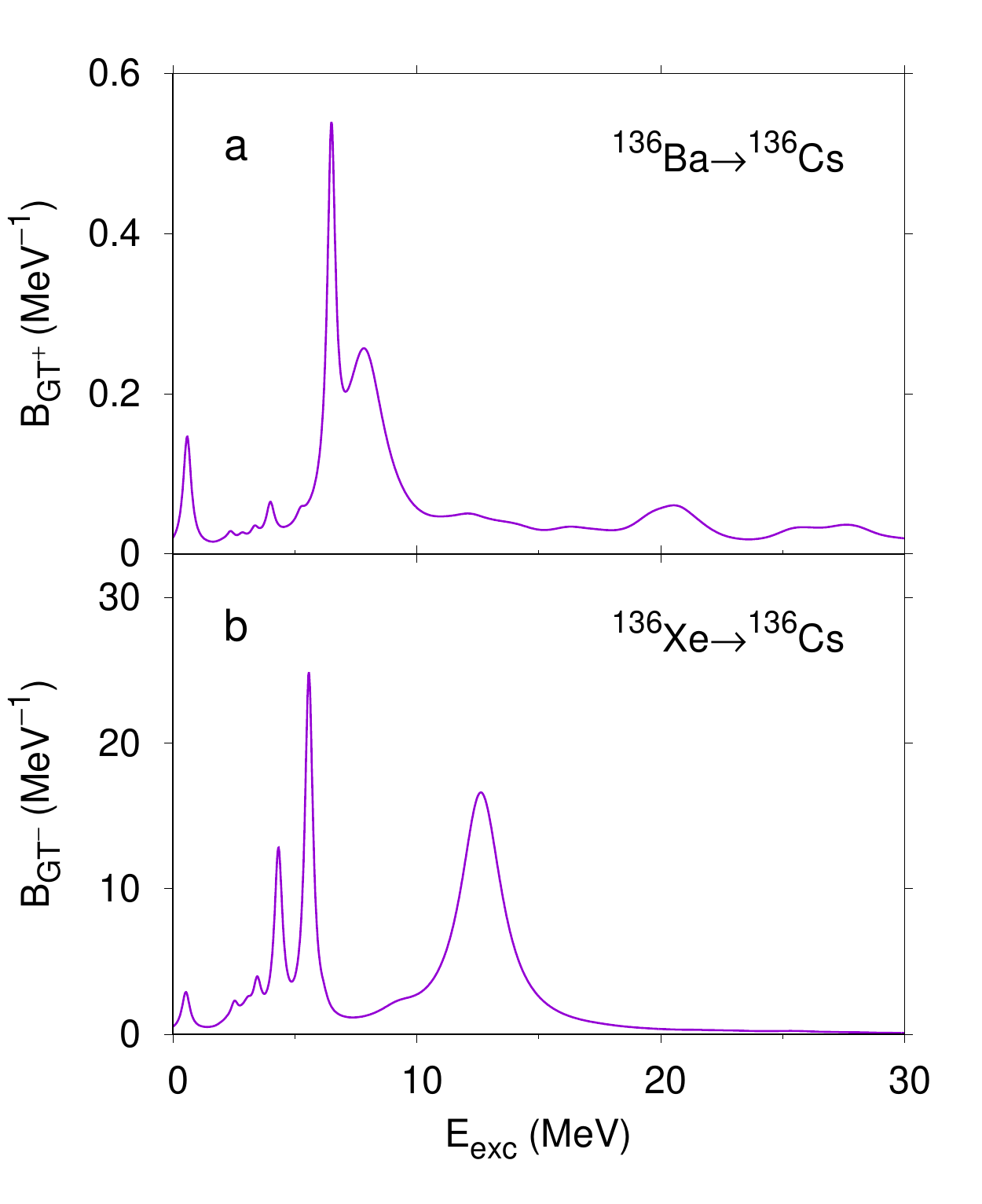}
\vspace{-10pt}
\caption{ \protect \label{fig:strfn_gt+-} \baselineskip=13pt 
GT$^+$ strength function $B_{\textrm{GT}^+}$  from $^{136}$Ba to $^{136}$Cs (a) and GT$^-$ strength function $B_{\textrm{GT}^-}$ from $^{136}$Xe to $^{136}$Cs (b) as functions of $E_\textrm{exc}$ of $^{136}$Cs by my normal QRPA calculation. }
\end{figure}

The above discussion is on the NME from the initial state to the intermediate state with the sign change. 
If the NME from this intermediate to the final state keeps the sign under the enhancement of $\chi$, the component of $M^{(2\nu)}_\textrm{GT}$ with this intermediate state changes its sign. 
%
The NME of the single-GT transition can be written
\begin{eqnarray}
&&\langle B | \bm{\sigma}\tau^- | I \rangle \approx \bm{t}_0 \psi_0 + \bm{t}_1 \psi_1 + \cdots, \label{eq:GTNMEapprox-1} \\
&&\langle F | \bm{\sigma}\tau^- | B \rangle 
=\langle B | \bm{\sigma}\tau^+ | F \rangle
\approx \bm{t}_0^\prime \psi_0^\prime + \bm{t}_1^\prime \psi_1^\prime + \cdots, \label{eq:GTNMEapprox-2}
\end{eqnarray}
where $\bm{t}_k$ $(k=0, 1,\cdots)$ denotes the single-particle matrix elements of $\bm{\sigma}\tau^-$ ($k$ is the abbreviation of $\mu_k i_k$), and $\bm{t}_k^\prime$ is that for the GT$^+$ ($p$ $\rightarrow$ $n$) transition operator $\bm{\sigma}\tau^+$ = $(\bm{\sigma}\tau^-)^\dagger$. 
The definition of $\psi_k^\prime$ is analogous to that of $\psi_k$ but for the GT$^+$ transition from $|F \rangle$ to $|B \rangle$ so that the $k$ of $\psi_k^\prime$ is different from the $\mu_k i_k$ of $\psi_k$.  
$|B_I\rangle$ $\approx$ $|B_F\rangle$ $\approx$ $|B\rangle $ is assumed for simplifying the discussion. 
Let $k$ = 0 of $\psi_k$ correspond to the single-particle states of the spin-orbit partner. 
Now, the GT strength function and the GT double-$\beta$ NME share the GT NME. 
The GT$^+$ strength function of Fig.~\ref{fig:strfn_gt+-} does not have a peak  at the peak energy of the GT$^-$ giant resonance (12.5 MeV). This implies that the GT$^+$ transition strength at this energy approximately does not depend on the component of $|B\rangle$ causing the GT$^-$ peak. If $|F\rangle$ is approximated to be a single-Slater state, the GT$^+$ transition from this $|F\rangle$ cannot create that peak-creating component of $|B\rangle$. 
Namely, the GT NME (\ref{eq:GTNMEapprox-2}) does not have a major component corresponding to $\bm{t}_0 \psi_0$.
The major components of this NME do not change their signs because the associated 
$\epsilon_\mu - \epsilon_i$ are not close to the $E$ corresponding to the GT$^-$  peak [see Eq.~(\ref{eq:amp})]. 
Therefore, the possibility of the sign change of $\langle F|\bm{\sigma}\tau^-|B\rangle$ is lower than that of $\langle B|\bm{\sigma}\tau^-|I\rangle$. 
Actually, the RPA solution of $|B_F\rangle$ has the peak-creating component of $|B\rangle$ only due to the ground-state correlations. 
%

The enhancement of the attractive interaction explains the reduction of many components of $M^{(2\nu)}_\textrm{GT}$ and is also a necessary condition for the sign inversion of the component with the GT giant resonance. 
Therefore, it is inferred that the essential cause of the discrepancy problem of the running sum is the difference in the strength of the attractive residual interaction. 
Another cause is that the shell model calculation in Fig.~\ref{fig:QRPA_SM} does not include the 0$g_{9/2}$ or 0$h_{9/2}$ orbitals. This is the reason why the running sum of that shell model calculation shows no  change around $E_\textrm{exc}$ = 12.5 MeV. 

\section{\label{sec:validity_interaction} Validity of the interaction}
Hence, it is crucial to examine the strength of the residual interactions. For the charge change phenomena, however, this is not straightforward. The calculation of the GT transition strength by the weak interaction needs an effective $g_A^\textrm{eff}$ with an uncertainty, and the equivalent transition strength obtained by the charge exchange reactions is accompanied by an uncertainty due to the contributions of operators other than the GT operator, e.g.,~the isovector spin monopole operator \cite{Yak09}. The Skyrme and the contact pairing interactions are used in my HFB and QRPA calculations. Besides the binding energy discussed above,  this approach can be also examined by the applications to the electric transitions. I use large single-particle spaces, so that an effective charge is not necessary. The validity of my approach has been proven systematically with SkM$^\ast$ for the energies and the electric transition strengths for the low-lying excited states of medium-heavy even-even nuclei \cite{Ter08}. For the like-particle pairing interactions, the strength is determined so as to reproduce the experimental pairing gaps obtained by the three-point formula \cite{Boh69} as usual. The $pn$ pairing correlations are active on and around the $N=Z$ line ($N$ is the neutron number,  and $Z$ is the proton number) in the nuclear chart and weakened for nuclei far from that line; see references in Ref.~\cite{Ter20}. For $^{136}$Xe, the $pn$ pairing correlations cannot be so strong as to make this nucleus close to the phase transition to the $pn$ pair condensation. My interaction strengths are not  insufficient. Thus, the nearly monotonic behavior of the running sum is reasonable. 

My mild interaction has a counter part that the $M^{(2\nu)}_\textrm{GT}$ is much larger than those of the QRPA calculation of Fig.~1, and my phenomenological $g^\textrm{eff}_A$ to reproduce the experimental half-life of the $2\nu\beta\beta$ decay is 0.49 \cite{Ter19}. This implies that large many-body effects are necessary for the microscopic derivation of the effective GT operator, which is not yet achieved. 

Finally, I mention why the double-$\beta$ NME is so sensitive to the interaction. This is because the GT$^+$ transition is strongly hindered for neutron rich nuclei and constrained by the GT (Ikeda) sum rule \cite{Suh07}, which states that the summation of the GT$^-$ strengths $S_\textrm{GT}^-$ subtracted by that of the GT$^+$ strengths $S_\textrm{GT}^+$ from the common nucleus of $(Z,N)$ is equal to $3(N-Z)$. This sum rule can be satisfied well by using the sufficiently large single-particle spaces. $S_\textrm{GT}^-$ and $S_\textrm{GT}^+$ for $^{136}$Xe are 85.145 and 1.139, respectively, in my normal QRPA calculation, and the reference calculation with $G^\textrm{IS}_{pn}$ = $-$280.0 MeVfm$^3$ yields 
$S_\textrm{GT}^-$ = 84.725 and $S_\textrm{GT}^+$ = 0.718. 
The enhancement of $G^\textrm{IS}_{pn}$ reduces the $S_\textrm{GT}^+$ by 37 \%. 
One may think that if the experimental GT$^+$ strength functions by the weak probe are reproduced, the reliability of the prediction of the $0\nu\beta\beta$ NME would be high, however, there is the $g_A^\textrm{eff}$ problem. Recently, the muon capture rate was calculated by two groups \cite{Jok19,Sim20}, but their results are rather different. 

\section{\label{sec:Summary} Summary}
I have examined several possibilities causing the discrepancy problem of the calculated running sums for the NME of the $2\nu\beta\beta$ decay of $^{136}$Xe. This is a serious problem casting doubt on the reliability of the double-$\beta$ calculations, thus, affecting neutrino physics, and it has been a mystery for four years since the problem was presented. My calculation shows a nearly monotonic increasing behavior, but there are results with large increase and decrease by other groups. My test calculations could not reproduce quantitatively those non-monotonic behaviors, but a weak similarity was obtained by strongly enhancing the isoscalar $pn$ pairing interaction. I have discussed the sign of the GT transition matrix elements analytically, and it has turned out that the enhancement of the attractive interaction is a necessary condition to cause the negative NME component. The origin of the sign inversion of the transition matrix element is that the main component of the GT giant resonance moves from a solution to another one. 
The analytical discussion complements the realistic calculations because these calculations do not explain the cause of the negative NME component. 
I conclude from these discussions that the essential cause of the discrepancy problem is the large strengths of  the attractive interactions. 
Concerning my approach, the strengths of the interactions are sufficient. Thus, it is also concluded that the nearly monotonic running sum is reasonable. 

\begin{acknowledgments}  
I am grateful to Dr.~Men\'{e}ndez for giving me the data of Fig.~1 and the fruitful discussion. 
This study was supported by European Regional Development Fund, Project ``Engineering applications of microworld physics" (No.~CZ.02.1.01/0.0/0.0/16\_019/0000766). The numerical calculations of this article were performed using the computer Karolina of IT4Innovations, National Supercomputer Center  (open-26-57) supported by the Ministry of Education, Youth and Sports of the Czech Republic through the e-INFRA CZ (ID:90140 and 90254), 
the computers of MetaCentrum (terasjun) supplied by the project "e-Infrastruktura CZ" (e-INFRA CZ
LM2018140 ) supported by the Ministry of Education, Youth and Sports of the Czech Republic,
the computer Cygnus at Center for Computational Sciences, University of Tsukuba through the Multidisciplinary Cooperative Research Program 2022 (NUCBETA), and
the computer Yukawa-21 at Computer Facility, Yukawa Institute for Theoretical Physics, Kyoto University (jun.terasaki). 
Parts of the calculations were also performed by the computer Oakforest-PACS of the Joint Center for Advanced High Performance Computing through the program of the High Performance Computing Infrastructure in the fiscal year 2021 (hp210001). 
\end{acknowledgments}

\appendix*
\section{\label{sec:appendix1} Illustative example}
\begin{figure}[t]
\includegraphics[width=0.5\columnwidth]{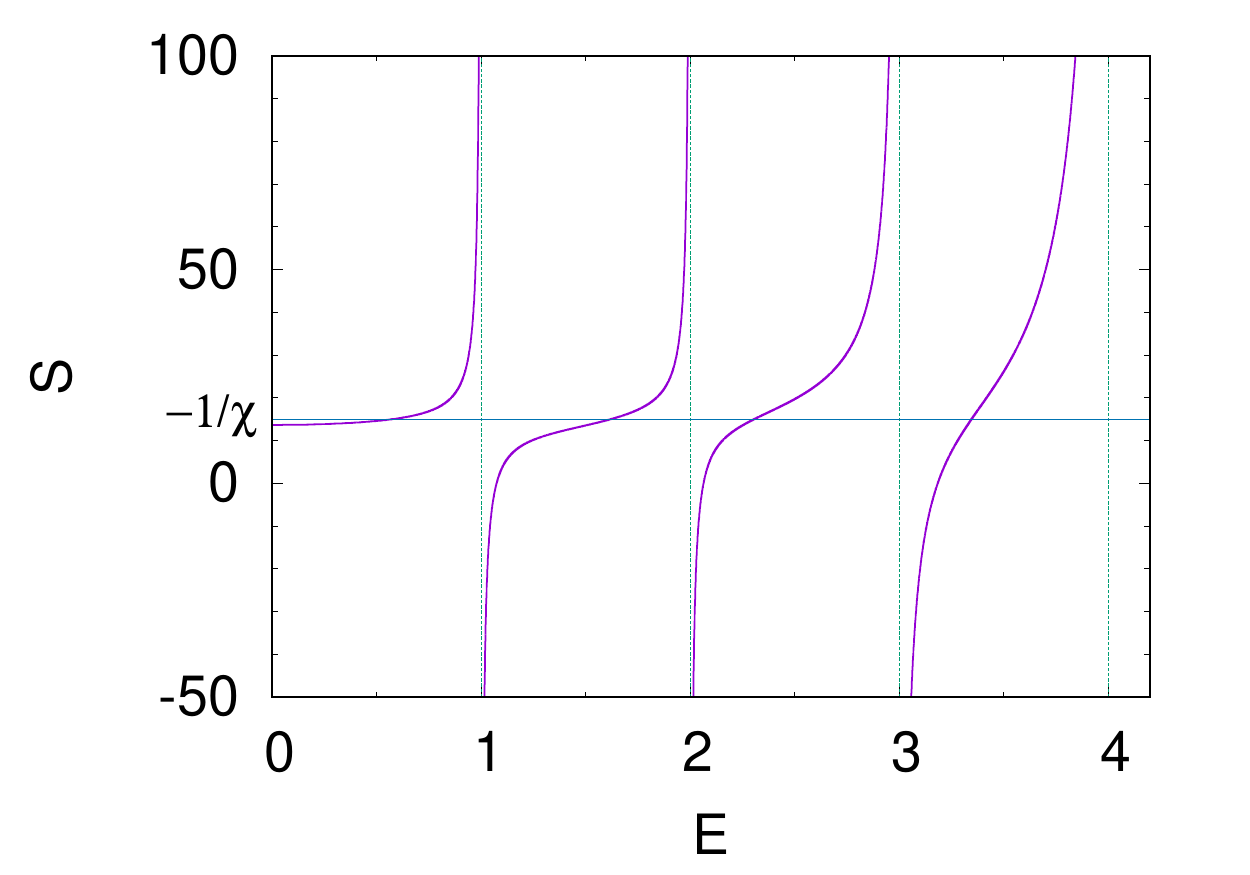}
\vspace{-10pt}
\caption{ \protect \label{fig:simple3} \baselineskip=13pt 
A simple example of $S(E)$ [Eq.~(\ref{eq:secular})] and $-1/\chi$ shown by horizontal line. The vertical lines show the unperturbed energies. $E$ and $\chi$ are dimensionless in this example. }
\end{figure}

Figure \ref{fig:simple3} shows a schematic example of $S(E)$ of Eq.~(\ref{eq:secular}). 
The crossing points of $S(E)$ and the horizontal line of $-1/\chi$ give the eigen energies of the RPA. 
$E$ =1, 2, 3, and 4 are the unperturbed energies $\epsilon_\mu-\epsilon_i$. I define the solution number in ascending order of the eigen energy. 
Suppose that $E$ = 3 is the unperturbed energy of the GT giant resonance, that is, $\mu i$ of Eq.~(\ref{eq:amp}) are the spin-orbit partners. 
If a small negative $\chi$ is introduced, the eigen energy nearest to $E$ = 3 is slightly lower than this $E$.  
The solution 3 is of the GT giant resonance, and  
the main amplitude $\psi_{\mu i}$ of this solution has the positive energy denominator. If  $\chi$ is enhanced, e.g.,~as the one shown in the figure, the eigen energy nearest to $E$ = 3 is higher than this $E$. Therefore, the GT giant resonance moves from the solution 3 to 4, and the sign of the energy denominator of the main $\psi_{\mu i}$ changes. This explains the origin of the sign change of 
the component of the double-$\beta$ NME 
in the enhancement of the interaction strength. 
If the normalization factor has different signs for the solutions 3 and 4, the sign of the NME from $|I\rangle$ to $|B\rangle$ does not change, and that from $|B\rangle$ to $|F\rangle$ changes. For the latter NME, see Sec.~\ref{sec:Analytical_discussion}. 

Actually, I chose parameters $C_{\mu i}$ to give rise to this sign change. Thus, the occurrence of the sign change depends on not only the interaction strength but also the operator part of the interaction. 
%
\end{document}